\documentclass{article}
\usepackage{textcomp}
\usepackage{hyperref}
\usepackage{geometry}
\usepackage{amsmath}
\usepackage{authblk}
\usepackage[utf8]{inputenc}
\usepackage{graphicx}

\title{Laboratory x-ray nano-computed tomography for biomedical research%\\
       }
\author[1,2]{Till Dreier}
\author[1]{Robin Krüger}
\author[3]{Gustaf Bernström}
\author[3,4,5]{Karin Tran-Lundmark}
\author[6]{Isabel Gonçalves}
\author[1,7]{Martin Bech}

\affil[1]{Lund University, Department of Medical Radiation Physics, Clinical Sciences Lund, Barngatan 4, 22242 Lund, Sweden}
\affil[2]{Excillum AB, Jan Stenbecks Torg 17, 16440 Kista, Sweden}
\affil[3]{Department of Experimental Medical Science, Faculty of Medicine, Lund University, Lund, Sweden}
\affil[4]{Wallenberg Center for Molecular Medicine, Lund University, Lund, Sweden}
\affil[5]{The Pediatric Heart Center, Skåne University Hospital, Lund, Sweden}
\affil[6]{Cardiology, Skåne University Hospital and Department of Clinical Sciences Malmö, Lund University, Lund, Sweden}
\affil[7]{LINXS Institute for Advanced X-ray and Neutron Science, Lund, Sweden}

\begin{document}

% punish hyphenation
\hyphenpenalty=10000
\emergencystretch=100em

\maketitle

\begin{abstract}
    High-resolution x-ray tomography is a common technique for biomedical research using synchrotron sources. With advancements in laboratory x-ray sources, an increasing number of experiments can be performed in the lab. In this paper, the design, implementation, and verification of a laboratory setup for x-ray nano-computed tomography is presented using a nano-focus x-ray source and high geometric magnification not requiring any optical elements. Comparing a scintillator-based detector to a photon counting detector shows a clear benefit of using photon counting detectors for these applications, where the flux of the x-ray source is limited and samples have low contrast. Sample contrast is enhanced using propagation-based phase contrast. The resolution of the system is verified using 2D resolution charts and using Fourier Ring Correlation on reconstructed CT slices. Evaluating noise and contrast highlights the benefits of photon counting detectors and the contrast improvement through phase contrast. The implemented setup is capable of reaching sub-micron resolution and satisfying contrast in biological samples, like paraffin embedded tissue.
\end{abstract}

\section{Introduction}
% intro to nano imaging (lab + synchrotron)
High-resolution x-ray tomography has become a widely used tool in biomedical research. While synchrotron imaging is the gold standard allowing for fast scans with high resolution and with excellent contrast through the use of propagation-based phase contrast imaging (PB-PCI) \cite{Snigirev1995} or x-ray holography \cite{Cloetens1999}, laboratory setups are becoming viable alternatives. With developments of brighter micro-focus sources \cite{Hemberg2003} and nano-focus sources \cite{Nachtrab2014,Muller2017a,Ferstl2018}, an increasing amount of biomedical imaging applications are becoming feasible in the laboratory. Investigations of tissue samples are often performed with resolution of around 1~\textmu m or with sub-micron resolution down to a few hundred nm. The setup presented in this paper can achieve sub-micron resolution in biological samples, such as tissue samples, with sufficient contrast to enable biomedical studies. \\

% methods to achieve nano resolution (small spots, optics, etc) 
There are a variety of methods to achieve high-resolution in x-ray tomography: incorporating visible light optics into a scintillator-based detector \cite{Stampanoni2006}, x-ray optics \cite{Niemann1976,Withers2007,Robinson2009} to high geometric magnifications using small x-ray spots \cite{Muller2017a,Fella2018,Eckermann2020}, and techniques like ptychography \cite{Rodenburg2007}. Using visible light optics to achieve high resolution is a common approach at synchrotron sources \cite{Stampanoni2006} and also in laboratory imaging \cite{Topperwien2016a,Dierks2022,Rashidi2020}. Alternatively, x-ray optics have been used to achieve even higher resolutions \cite{Fella2017,Jacobsen2019}. However, such optics-based approaches typically require monochromatic and highly coherent x-rays \cite{Niemann1976}. Lens-less approaches like ptychography \cite{Rodenburg2007,Thibault2008,Pfeiffer2018} can provide extremely high resolution not limited by the spot size given coherent illumination. Recent developments are working towards making ptychography available in the laboratory as well \cite{Batey2021}. \\

% High res in the lab 
% comparison/verification of nano-CT (Göttingen) between synchrotron and lab setup
% 	IS THERE A PUBLICATION ON ZONE PLATES FOR HIGHER X-RAY ENERGIES????? deepXscan?
While x-ray optics require quasi-monochromatic illumination and are only widely available for lower x-ray energies \cite{Muller2021} and magnifying detectors utilise low-efficiency scintillation detectors, high geometric magnification with a small x-ray spot can provide an alternative to achieve very high resolution in a laboratory setup \cite{Nachtrab2014,Fella2018}. Such setups are very similar to established micro-CT systems, but utilise transmission x-ray sources (as opposed to reflection-type x-ray sources), allowing us to place the sample significantly closer to the x-ray source (closer than 1~mm) resulting in very high magnification factors while retaining a compact system \cite{Fella2018}. This has clear advantages for higher energy applications as shown by Graetz et al \cite{Graetz2020} and Müller et al \cite{Muller2021,Mueller2022}. Studies on biomedical samples have also shown that laboratory nano-tomography can provide sufficient data quality for medical studies \cite{Ferstl2018,Eckermann2020,Romell2021}. \\

% Limiting factors (spot size, magnification, detector, etc) + a sentence on super-res (movable spot) with ref to Jonas
% The achievable resolution in a high magnification imaging system is limited by the size of the x-ray spot to half the spot size assuming a Gaussian spot shape. In an imaging, other factors such as penumbral blurring from the spot, blurring from the detector response function, and instabilities and drifts of the imaging system affect the achievable resolution \cite{Dreier2021}. Approaches like super-resolution have shown that the resolution can be improved through oversampling as long as the spot size is not the limiting factor \cite{Dreier2021,Dreier2020}. The same concept has been utilised by Graetz et al \cite{Graetz2020} for nano imaging. \\

% challenges to achieve nano resolution (source requirements, drifts, environment, etc)
% challenges with biomedical (or soft) samples
Building a laboratory nano-tomography imaging system comes with a variety of challenges and requirements to the equipment. Thermal and mechanical stability are a main concern, which includes the x-ray source, stages, detector, and the environment of the system. Another concern is the scan quality, i.e. the achievable contrast and scan times required for samples consisting of low-Z materials. The following section discusses how the requirements and parameters were taken into account to build the imaging system. \\

\section{Methodology}
\subsection{Requirements for biomedical nano-CT}
% contrast requirements
%   absorption difference paraffin/tissue
%   artefacts and effects
In low-density samples, such as soft tissue, high contrast is best achieved with low energy x-ray photons. With samples that are paraffin embedded and small enough to not absorb too much of the lower energy photons, good soft tissue contrast can be achieved using propagation-based phase contrast \cite{Snigirev1995,Wilkins1996}. Thus the sample size should be minimised, as well as the distance to the detector reducing absorption in air. Denser parts of the sample might lead to beam hardening artefacts. Conventional staining agents are typically very dense compared to tissue. Hence, no staining should be performed ideally, or appropriate contrast agents should be selected \cite{Dreier2022}. % Embedding a sample in a denser material, such as plastic, epoxy, or a solution containing water, will cause significant absorption resulting in reduced contrast. \\

The biological tissue samples used as test objects in this study were fixated and embedded in paraffin. Sample preparation of the atherosclerotic plaque is described in \cite{Truong2022}, whereas the procedure for the cow lung injected with dye is described in \cite{Dreier2022}.

%Short sample preparation section under ”Methodology”? Or enough to refer to the micro-CT publication? May be good to at least mention the dye injections?

% distances (magnification/p_eff vs pixel size)
High resolution is achieved through geometric magnification using a diverging cone beam. Accordingly, the effective pixel size $p_{\text{eff}} = P/M$, with $P$ as the physical pixel size of the detector, depends on the magnification $M = (R_1 + R_2)/R_1$, where $R_1$ is the source-sample distance, and $R_2$ is the sample-detector distance. To optimise phase contrast, the effective propagation distance needs to be considered, which scales with the magnification: $z_{\text{eff}} = R_2/M = R_1 R_2 / (R_1 + R_2)$. Evaluation and optimisation of phase fringes was performed by Dierks et al. \cite{Dierks2023}. \\

% drifts, vibrations, etc + Source requirements
Another requirement is a stable x-ray spot from the source, both in size and position. Drift or defocusing of the x-ray spot will lead to a loss of resolution. The whole imaging system is affected by thermal drift leading to slow movements of the equipment during the measurements, which can be corrected using alignment images, as described in Section \ref{ch:data_acquisiotn}. Vibrations are much more difficult to correct, particularly in laboratory setups with long exposure times, and should be avoided through careful design of the setup. However, vibrations cannot always be excluded completely and can be caused by instabilities of the equipment, i.e. precision of the motion axes, air flow, or other equipment nearby utilising a pump or cooling fan. Vibrations that can be resolved, i.e. when the frequency is sufficiently low while the exposure time of the detector is sufficiently short, can be corrected using algorithms like joint reprojection \cite{Gursoy2017}, phase symmetry \cite{Pande2022}, or distributed optimisation \cite{Nikitin2021}.  \\

% phase contrast and distances
To improve contrast in low density samples, such as soft tissue, propagation-based phase contrast \cite{Snigirev1995,Bravin2013} is often utilised. Propagation-based phase contrast considers illumination of a sample with a wave, which is perturbed by interfaces between materials with different refraction indices. Given a sufficiently large propagation distance, phase shifts are detected as intensity variations on the detector. Considering illumination with a magnifying cone beam, as in any optics-free laboratory x-ray imaging system, the geometric magnification has to be taken into account to find the effective propagation distance $z_{\text{eff}} = R_2/M$. Since reducing $M$ up to the point where $R_1=R_2$ will increase $z_{\text{eff}}$ while reducing the effective pixel size $p_{\text{eff}} = P/M$ limiting the achievable resolution. Thus, optimising $z_{\text{eff}}$ and $p_{\text{eff}}$ might require to increase $R_1$ and $R_2$ and thus absorption in air, resulting in longer scan times increasing effects from thermal drifts. \\

A variety of phase-retrieval algorithms have been discussed in detail by Burvall et al \cite{Burvall2011,Burvall2013}, a general summary for x-ray imaging has been provided by Nugent et al \cite{Nugent2007}, and practical considerations have been described by Gureyev et al \cite{Gureyev2009}. A common approach is single material phase-retrieval \cite{Paganin2002}. Another algorithm working well for laboratory data is the Bronnikov Aided Correction (BAC) \cite{DeWitte2007}. \\

% Detector selection
%   params to consider
Since resolution is achieved through a small x-ray spot combined with high geometric magnification, $R_1$ and $R_2$ have to be selected to yield a sufficiently small effective pixel size $p_{\text{eff}}$. Further, it is beneficial to reduce the total length of the setup to minimise absorption in air while maximising photons per area on the sample. Hence, there are two detector parameters to consider: physical pixel size and detection efficiency. \\

%   types of detectors
Common x-ray cameras are sCMOS (scientific Complementary Metal–Oxide–Semiconductor) detectors. These detectors use a scintillator to convert incoming x-rays into visible light, which is then detected using a sCMOS camera sensor. One of their main advantages is that small pixels (down to a few \textmu m) can be achieved. However, the point-spread function (PSF) of the camera sensor is typically a few pixels large, i.e. visible light photons from a single scintillation event are detected in several neighbouring pixels resulting in blurring of the image. Additionally, the scintillator material and thickness also might cause blurring and affect the detection efficiency.

%   Scintillator detectors (energy efficiency, noise, PSF, pixel size, etc)
To achieve high resolution at low x-ray energies, the scintillator has to be as thin as possible to minimise blurring while being as efficient as possible for the relevant energies. A material such as Gadolinium Oxysulfide (Gadox) can be produced very thin, down to a few \textmu m, with decent efficiency at around 10~keV allowing high resolution at lower energies. %, while materials like Caesium Iodine (CsI) typically is much thicker, provides a much higher sensitivity, and is suitable for high energies $>$50~keV \cite{Cha2010}. 
Besides blurring from the scintillator and PSF of the detector, noise has to be considered as well. Readout noise and electronic noise are a significant concern for photon-limited applications. \\

%   Photon counting detectors (energy efficiency, noise, PSF, pixel size, ...)
As an alternative to scintilllation based cameras, photon counting detectors provide a significantly smaller PSF close to a single pixel \cite{Rossi2006}, lower noise, and higher efficiency at the expense of larger pixels making them a viable alternative for biomedical and high resolution imaging \cite{Bech2008,Flenner2023}. \\
Photon counting detectors detect x-ray photons directly in a semiconductor sensor. Every pixel has its own dedicated electronics allowing to suppress noise and reduce cross talk between pixels yielding an almost box-like PSF covering a single pixel \cite{Johnson2014}. Using a silicon sensor yields almost 100~\% efficiency for 10~keV photons dropping with increasing photon energy \cite{Donath2013,Donath2023}. A comparison between the theoretical and practical efficiency of a Pilatus detector \cite{Henrich2009} has been shown by Donath et al \cite{Donath2013}. These properties make photon counting detectors a viable choice for imaging of low density samples especially with limited flux as demonstrated by Scholz et al \cite{Scholz2020} and Dudak et al \cite{Dudak2022c} or for high resolution as shown by Flenner et al \cite{Flenner2023}. However, their main limitation for this particular application is the pixel size (75~\textmu m for the Eiger detector implemented in our work), which is significantly larger than those of sCMOS cameras. It should be noted, however, that pixel size cannot be directly compared between scintillator-based and photon-counting detector due to their different PSFs. Using a photon-counting detector, equivalent resolution can be achieved with a larger $p_{\text{eff}}$, thus reducing the required distance, compared to a sCMOS detector. \\

\subsection{Setup design}
% FIGURE 1: setup sketch
\begin{figure}[t]
    \centering
    \includegraphics[width=\textwidth]{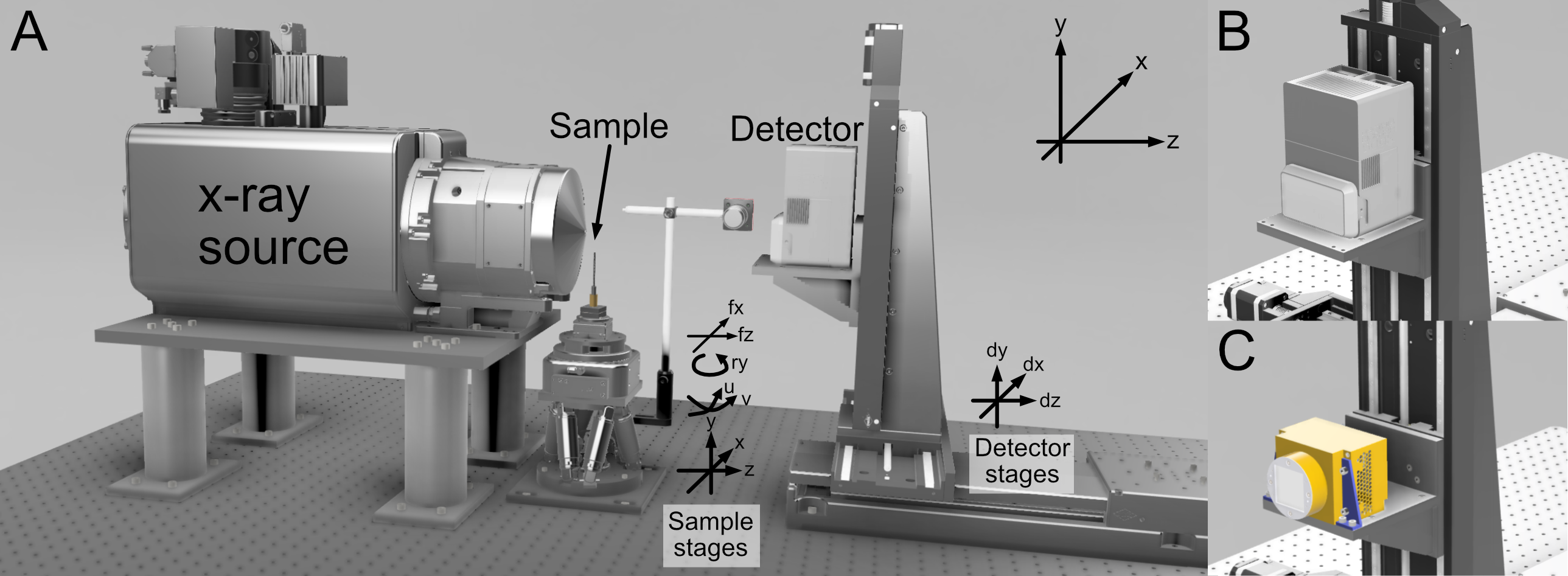}
    \caption{(A) 3D model of the nano-CT setup. The sample holder consists of a hexapod with 6 axis, of which \textit{x, y, z, u, v} are used. Positioning of the detector is also motorised along the three main axes (\textit{dx, dy, dx}). (B) Zoom in on the detector mount with the photon counting Eiger 2R 500K mounted. (C) Zoom in on the detector mount with the scintillator-based Photonic Science GSense 16M mounted.}
    \label{fig:setup}
\end{figure}

% coordinate system
In the setup (Figure \ref{fig:setup}A), the following coordinate system is used: \textit{z} represents the horizontal axis along the beam from the source towards the detector, \textit{x} represents the perpendicular horizontal axis, and \textit{y} represents the vertical axis. Tilt is represented by \textit{u} perpendicular to the beam and \textit{v} along beam direction. Rotation around the $y$ axis is referred to as \textit{ry}. Another coordinate system on top of the rotation axis have axes referred to as \textit{fx} and \textit{fz}, coinciding with the $x$ and $z$ axes respectively with the rotation \textit{ry} at 0 degrees. \\ 

% x-ray source
The setup utilises a NanoTube N2 60 kV nano-focus x-ray source (Excillum AB, Sweden), which can achieve spot sizes from 1.2~\textmu m down to 300~nm. The acceleration voltage can be set between 40 and 60~kV. The source utilises a patterned tungsten on diamond transmission target, which is used to calibrate the desired spot size, internally verify the spot size, spot position on the target, and sets the maximum safe power automatically. The main energy emitted is around 8~keV (Tungsten $L_\alpha$), matching well to the efficiency of the detectors. \\

% sample stages
A 9-axis sample holder is constructed from a H-811 hexapod (Physik Intrumente GmbH, Germany) allowing translation in \textit{x, y, z} as well as tilting along and perpendicular to the beam direction (\textit{u, v}). The hexapod is used for coarse positioning of the sample, i.e. height and centering of the rotation axis, and tilt correction of the rotation axis with a maximum range of $\pm$16~mm in \textit{x,z} and $\pm$6.5~mm in \textit{y} direction. On top, a RT100S air bearing rotation stage (LAB Motion Systems, Belgium) is mounted providing rotation around the \textit{y} axis. The stage was chosen for its positional accuracy, stability, and position repeatability. For fine positioning on top of the rotation axis, a U723 dual-axis piezo stage (Physik Instrumente GmbH, Germany) with ±11~mm range in \textit{x, z} direction and a positional accuracy of 10~nm, allowing centering of the sample or feature on the rotation axis. The distance from the x-ray spot to the rotation axis of the sample (source-object distance) is particularly critical for the achievable resolution. Slight differences can have significant impact on $p_{\text{eff}}$. \\

% detector stages
The 3-axis detector positioning system consists of three linear stages (Owis GmbH, Germany) allowing for detector alignment in \textit{x, y, z} direction. A long-range \textit{dz} stage with 0.6~m range was chosen to be able to select a wide range of magnifications. Using the \textit{dx, dy} stages, the detector can be aligned to the optical axis of the system. The two detectors are shown in Figure \ref{fig:setup}B--C.  \\ 

% sample mounting
The top of the sample holder is detachable using a KB25 kinematic base (ThorLabs Inc., USA). Different holders were designed to allow a variety of sample types and sizes. As primary mounting approach polyimide tubes were selected, which can be glued into a variety of 3D printed pins. Larger or less fragile sample can be glued directly to carbon fibre tubes (2~mm outer diameter, 1~mm inner diameter) using wax (paraffin wax or beeswax) or superglue. For small polyimide tubes, self-aligning 3D printed pins with a conical base \cite{Holler2017} are used. Larger polyimide tubes and carbon fibre tubes are glued into 3D printed 6~mm diameter pins, which are attached to the kinematic base on top of the piezo stage using a collet knob. \\

% other equipment
To avoid collisions of the sample with the source, a Zelux 1.6 MP CMOS high-resolution camera (ThorLabs Inc., USA) with a MVL50M23 lens (ThorLabs Inc., USA) is placed to the side monitoring the tip of the source. For a general overview, a M5055 PTZ (pan, tilt, zoom) camera is placed above the setup (AXIS Communications AB, Sweden). \\

% detectors
%   Eiger
%   Photonic Science
Two detectors are available in the setup, a photon counting Eiger 2R 500K (Dectris Ltd., Switzerland) and the sCMOS GSense 16M (Photonic Science Ltd., UK) with 9~\textmu m pixel size, a 4096$\times$4096 pixel sensor, and a 17.5~\textmu m Gadox scintillator. The Eiger 2R 500K detector  has a 450 \textmu m Silicon sensor with a pixel size of 75 \textmu m, and 1028 $\times$ 514 pixels. 

% environmental requirements
To achieve sufficient stability, a vibration-free and thermally stable environment is required. To reduce vibrations, e.g. from surrounding equipment, the setup was built on an optical table with air-cushioned feet. The experimental environment needs to be able to achieve thermal equilibrium to avoid drifts of the equipment itself through thermal expansion or contraction. Airflow needs to be minimised to avoid instabilities of, particularly, the sample. Aluminium has a fairly high thermal expansion coefficient compared to other materials, i.e. it causes larger drifts when the temperature changes. It should also be noted that a lot of equipment is built from Aluminium, thus thermal stability needs to be taken into account when selecting components. \\

\subsection{Data acquisition}
\label{ch:data_acquisiotn}
% interfaces & control
% All equipment is connected to a local network via Ethernet and interfaced via a custom server application using the command interfaces of the controllers. All stage controllers provide a simple text-based interface allowing to send text commands via Ethernet or USB. The server application is used to unify the commands, i.e. the same commands are available for all stages. Further, this allows to control the equipment from multiple applications. The two detectors are interfaced directly via Python. Both detectors have their own dedicated readout computers. The Eiger is interfaced using an API based on HTTP get and set requests and allows to acquire images into compressed HDF5 files, receive them as a compressed stream, or as TIF images. The GSense camera comes with a GUI, which is used for alignment and finding of parameters, and a server, which is interfaced by text-based commands giving access to all settings and allowing to grab or write TIF images directly to a specified location. Measurements are executed using a Python script handling the configuration and measurement after the user performed sample alignment. \\

% alignment scheme
To align a sample for a scan, first the Centre-of-Rotation (CoR) is centred on the detector. Secondly, the sample or feature of interest is centred on the detector using the precise axes on top of the rotation axis. At 0 degrees sample rotation, the \textit{fx} axis is used, at 90 degrees sample rotation the \textit{fz} axis is used. \\

% acquisition scheme
Before starting the scan, the source is calibrated to the desired spot size and acceleration voltage. Then, the selected acquisition parameters are entered into a Python script, which will execute the scan. A measurement typically consists of flat-fields, a pre-alignment scan, the actual CT scan, and a post-alignment scan. With the Eiger detector, it is sufficient to acquire flat-fields before the scan, both, the silicon sensor and emission from the source, are sufficiently stable. For the GSense detector, dark-field and flat-field images are acquired before the scan using the detector software. To later correct for drifts during the scan, pre- and post-alignment images are acquired, i.e. the sample is rotated in e.g. 10 or 20 degree steps, where at each step a projection is acquired. \\

\subsection{Data processing}
% FIGURE 2: PROCESSING FLOW CHART
% \begin{figure}
%     \centering
%     \includegraphics[width=\textwidth]{Processing_Flowchart_Dummy.png}
%     \caption{Flow chart detailing the processing steps applied to acquired projections, where blue steps are always executed and green steps are optional. FIRST COR EXECUTED BEFORE RING FILTERING!!!!}
%     \label{fig:processing_flow_chart}
% \end{figure}

% outliers and flat-field
The acquired projections, flat-field images, and alignment images are loaded and broken, noisy, or outlier pixels are removed. The Eiger detector has an internal mask, setting bad pixels to a fixed value. Other outliers, or outliers in images acquired with the GSense detector, can be identified by, e.g. thresholding pixels above or below a certain value to consider them noisy or dead, or by identifying pixels that deviate a certain number of standard deviations from the image mean. To replace the identified bad pixels, values from a median filtered version of the same image are inserted. \\

% CoR and ring filtering
The Centre-of-Rotation (CoR) is found by performing multiple reconstructions of the centre slice, using a fan beam reconstruction, with varying shifts along the $x$ axis, which are then evaluated visually. This approach has proven to be the most reliable to find the CoR. With a first estimate for the CoR, ring filtering can be performed. A slice is reconstructed with different ring filters and parameters. Several filters are implemented as described in \cite{Vo2018,Munch2009}. After selecting a ring filter and fine tuning its parameters, the filter is applied to all projections.  \\

% drift correction
Following the ring filter, drift correction can be performed. It is necessary to first perform ring filtering since these filters work by identifying and removing stripes from the sinograms, which may be deformed by applying the drift correction. A cross correlation \cite{Guizar-Sicairos2008} between the alignment images and the projections is performed yielding shifts in $x$ and $y$ direction. To obtain shifts for each projection, linear interpolation is performed. Corrections corresponding to these shifts are then applied to all projections. \\

% phase retrieval
If necessary, either single material \cite{Paganin2002} or BAC \cite{DeWitte2007} phase-retrieval is applied. Besides the geometry of the scan, the algorithm uses an estimate of the mean photon energy and estimations of $\delta$ and $\beta$ of the complex refractive index $n = 1 - \delta + i\beta$. The effect on a single projection is estimated, which is inspected visually for removal of phase fringes at material borders and blurring caused by the algorithm. Following, an unsharp mask \cite{Sheppard2004,Paganin2020} can be applied to compensate some of the blurring caused by phase retrieval. With parameters selected, all projections are processed. \\

% reconstruction
%   tilt u + v correction
Reconstructions are performed using a GPU implementation of the Feldkamp-Davis-Kress (FDK) algorithm \cite{Feldkamp1984}, an efficient implementation of filtered back-projection for cone beam geometry, via the ASTRA toolbox \cite{VanAarle2015,VanAarle2016}. Fine tuning of shifts and tilts utilises single slice FDK reconstructions. To correct tilts of the rotation axis perpendicular (\textit{u}) and along the beam (\textit{v}), single slices for a range of values are reconstructed using the FDK algorithm and inspected visually. If a drift correction was performed or if a tilt correction is necessary, the CoR needs to be fine-tuned using the same approach. With all shifts and tilts resolved, the full volume is reconstructed.  \\

\subsection{Evaluation of resolution and image quality}
% assess resolution in 2D
Measuring the performance of an imaging system in 2D using resolution charts is a standard approach. These charts consist of differently sized and shaped metal structures (often Tungsten or Gold) on a thin membrane. Thus, their contrast is comparably high particularly when compared to biomedical samples. Commonly, line patterns are used to determine the achievable resolution. The charts used in this paper were placed approximately 200~\textmu m from the x-ray source. \\

% Assess resolution in 3D
To evaluate the resolution of reconstructed slices, Fourier Ring Correlation (FRC) is used \cite{VanHeel1987}. Compared to obtaining an Edge Spread Function (ESF) and Modulation Transfer Function (MTF) by fitting a sharp edge as described by slanted-edge method \cite{Estribeau2004} and previously used in \cite{Dreier2021}, the FRC takes a full slice into account resulting in a more general resolution estimation. To calculate the FRC, the acquired projections are split and two identical reconstructions are performed. Before correlation, shifts, rotation, and scaling of the selected slices are aligned with sub-pixel precision. The parameters are found by matching features in both slices using an Oriented FAST and Rotated BRIEF (ORB) classifier \cite{Rublee2011} and removing outliers with an iterative Random Sample Consensus (RANSAC) algorithm \cite{Fischler1981}. The resulting FRC curve is compared to a threshold criterion \cite{VanHeel2005}. Since the input data is split, the half-bit criterion is an appropriate measure. The last intersection of the FRC curve with the criterion yields the resolution limit of the reconstructed slice. \\

% Scan params
The resolution was evaluated using three different samples: a piece of cork (roughly 0.5~mm in size), a cylindrical  punch fragment from paraffin embedded bovine lung tissue (0.5~mm in diameter), and a cylindrical  punch from paraffin embedded human atherosclerotic plaque tissue (0.5~mm in diameter). Scans were performed with both detectors and the scan parameters are summarised in Table \ref{tab:scanparams}. 

\begin{table}
    \centering
    \caption{CT scan parameters}
    \begin{tabular}{|l|c|c|c|c|c|c|}
        \hline
                                & Cork              & Cork              & Lung          & Lung          & Plaque         & Plaque \\
                                & Eiger             & GSense            & Eiger         & GSense        & Eiger          & GSense \\
        \hline
         Projections            & 2000              & 3600              & 3200          & 2800          & 1800           & 3200 \\
         Exposure time          & 2~s               & 6~s               & 20~s          & 16~s          & 20~s           & 7.5~s\\
         Source-Object dist.    & 9.2~mm            & 3.47~mm           & 3.54~mm       & 4.25~mm       & 1.89~mm        & 1.65~mm\\
         Source-Detector dist.  & 428.1~mm          & 100.15~mm         & 388.1~mm      & 238.15~mm     & 237.2~mm       & 99.15~mm\\
         $p_{\text{eff}}$       & 1.61~\textmu m    & 623~nm            & 683~nm        & 321~nm        & 599~nm         & 298~nm\\
         X-ray spot size        & 1.2~\textmu m     & 600~nm            & 600~nm        & 400~nm        & 800~nm         & 900~nm\\
         Voltage                & 60~kV             & 60~kV             & 60~kV         & 60~kV         & 60~kV          & 60~kV\\
         \hline
    \end{tabular}
    \label{tab:scanparams}
\end{table}

% assess contrast and noise
To assess the image quality, the Signal-to-Noise ratio (SNR) and Contrast-to-Noise ratio (CNR) are calculated. Both methods work without a reference image and instead utilise a region on the sample and on the background area of the image. The SNR is defined as: 
\begin{equation}
    \text{SNR} = 20 \times \log \left( \frac{s}{n_{\text{RMS}}} \right),
    \label{eq:snr}
\end{equation}
\noindent where $s$ is the mean value of a region on the sample and $n_{\text{RMS}}$ is the root mean square of a region containing only background. The CNR is defined as: 
\begin{equation}
    \text{CNR} = \frac{|s-n|}{\sigma_n},
    \label{eq:cnr}
\end{equation}
\noindent where $s$ is the mean value of a region containing the sample, $n$ is the mean value of a region containing only background and $\sigma_n$ is the standard deviation of the region containing only background as previously described \cite{Bech2008,Dreier2020}. \\

\section{Results and discussion}
\subsection{Mechanical stability}
\begin{figure}
    \centering
    \includegraphics[width=\textwidth]{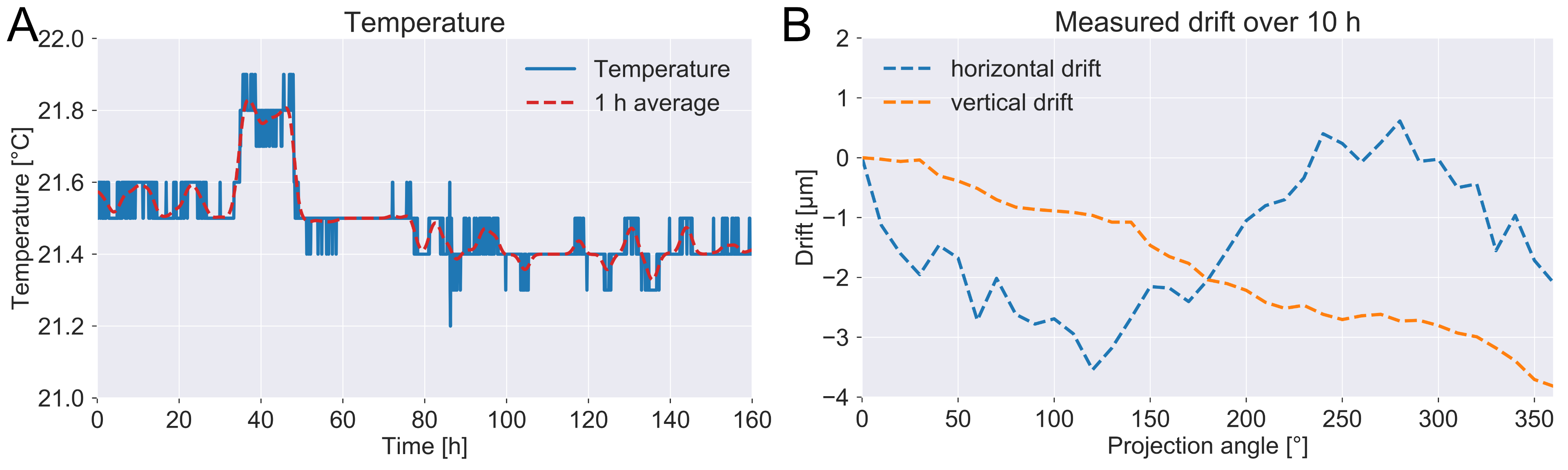}
    \caption{(A) Measured temperature in the experimental hutch over 160~h with an average temperature of 21.5~$^\circ$C and a standard deviation of $\pm$0.12 $^\circ$C. (B) Sample drift measured using alignment images of a 10~h CT scan with $p_{\text{eff}}$ = 626~nm.}
    \label{fig:stability}
\end{figure}

% temperature and drift
The temperature in the experimental hutch is monitored with a simple temperature logger and one data point per 5~s is acquired. Figure \ref{fig:stability}A shows a temperature measurement over 160~h with an average temperature of 21.5~$^\circ$C and standard deviation of $\pm$0.12~$^\circ$C. Accessing the hutch will cause a change in temperature. Hence, access should be limited in time and frequency to limit the time required to reach thermal equilibrium again. Figure \ref{fig:stability}B shows the measured drift during a scan measured using alignment images acquired before and after a scan. Slow drifts in the range of a few \textmu m could be measured, which can typically be corrected without issues. \\

\subsection{Resolution}
\begin{figure}
    \centering
    \includegraphics[width=\textwidth]{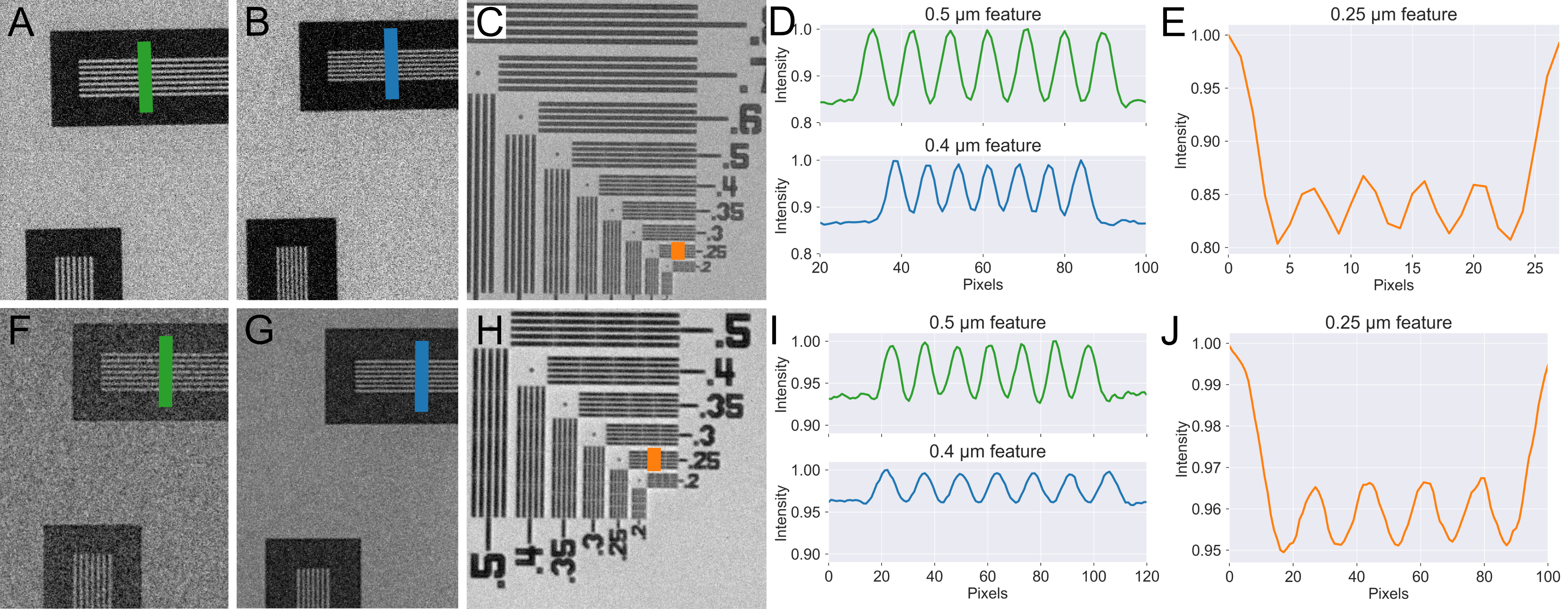}
        \caption{Measured resolution of the system using resolution charts. (A) 0.5~\textmu m feature on a JIMA chart imaged with the Eiger detector. (B) 0.4~\textmu m feature on a JIMA chart imaged with the Eiger detector. (C) Line patterns on a XRnanotech chart imaged with the Eiger detector. (D) Profile plots of the patterns in (A) and (B) as marked in their respective colour. (E) Profile plot of the smallest resolvable line pattern in (C) as marked in orange. (F) 0.5~\textmu m feature on a JIMA chart imaged with the GSense detector. (G) 0.4~\textmu m feature on a JIMA chart imaged with the GSense detector. (H) Line patterns on a XRnanotech chart imaged with the GSense detector. (I) Profile plots of the patterns in (F) and (G) as marked in their respective colour. (J) Profile plot of the smallest resolvable line pattern in (H) as marked in orange.}
    \label{fig:resolution_charts}
\end{figure}

% resolution charts
The 2D performance of the system was evaluated using line pattern on a JIMA RT RC-02 chart and a resolution chart from XRnanotech (Villigen, Switzerland). The smallest available features on the JIMA chart are 0.5~\textmu m and 0.4~\textmu m, which could be resolved by both detectors, as shown for the Eiger detector in Figure \ref{fig:resolution_charts}A,B,D and for the GSense detector in Figure \ref{fig:resolution_charts}F,G,J. On the XRnanotech chart, the 0.25~\textmu m feature could be resolved, as shown for the Eiger detector in Figure \ref{fig:resolution_charts}C,E and the GSense detector in Figure \ref{fig:resolution_charts}H,K. Profile plots in horizontal and vertical direction over multiple features are shown in Supplementary Figure \ref{fig:S2}. \\

% Figure: RESOLUTION CHARTS

\begin{figure}
    \centering
    \includegraphics[width=\textwidth]{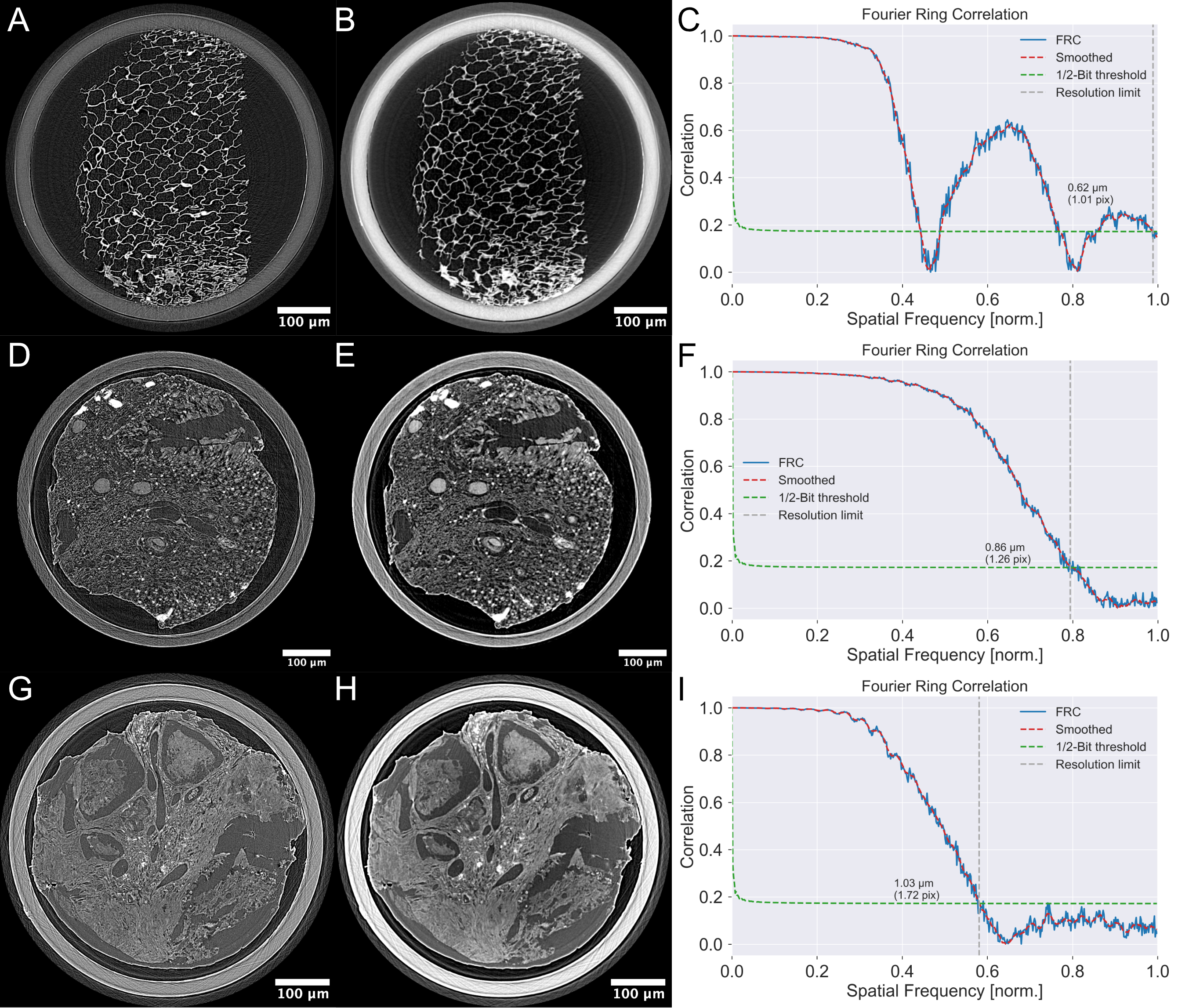}
        \caption{CT resolution using the Eiger detector. (A) Reconstructed slice with absorption contrast of a 0.5~mm piece of cork with a voxel size of 0.616~\textmu m. (B) Reconstructed slice with phase-retrieval applied. (C) Resolution determined by Fourier Ring Correlation. (D) Reconstructed slice with absorption contrast of a 0.5~mm punch fragment from a cow lung with a voxel size of 681~nm. (E) Reconstructed slice with phase-retrieval applied. (F) Resolution determined by Fourier Ring Correlation. (G) Reconstructed slice with absorption contrast of a 0.5~mm punch fragment from an atherosclerotic plaque sample. (H) Reconstructed slice with phase-retrieval applied. (I) Resolution determined by Fourier Ring Correlation.}
    \label{fig:eiger_resolution}
\end{figure}

% 3D resolution (EIGER)
Performance of the system in 3D was evaluated using FRC on three test samples: (i) a 0.5~mm piece of cork, (ii) a 0.5~mm punch fragment from a paraffin embedded cow lung (previously scanned with our micro-CT \cite{Dreier2022}), and (iii) a 0.5~mm fragment punched from a human atherosclerotic plaque (previously scanned with synchrotron micro-CT \cite{Truong2022}). The resolution is only measured on slices without phase-retrieval applied since phase-retrieval may corrupt the higher frequencies \cite{Riedel2023}. FRC of the lung punch fragment (Figure \ref{fig:eiger_resolution}F) shows that the achieved resolution is close to $p_{\text{eff}}$ at 1.26$\times p_{\text{eff}}$, while for the plaque punch fragment (Figure \ref{fig:eiger_resolution}I) a resolution of 1.72$\times p_{\text{eff}}$ could be achieved. The cork sample shows a resolutions of 1.01$\times p_{\text{eff}}$. Variations may be caused by the precision of the geometry alignment (CoR, tilt) and precision of the drift correction. \\

% Figure: EIGER FRC
% TO WRITE: almost no effect on resolution from phase retrieval, worse resolution in cork due to larger distances.

\begin{figure}
    \centering
    \includegraphics[width=\textwidth]{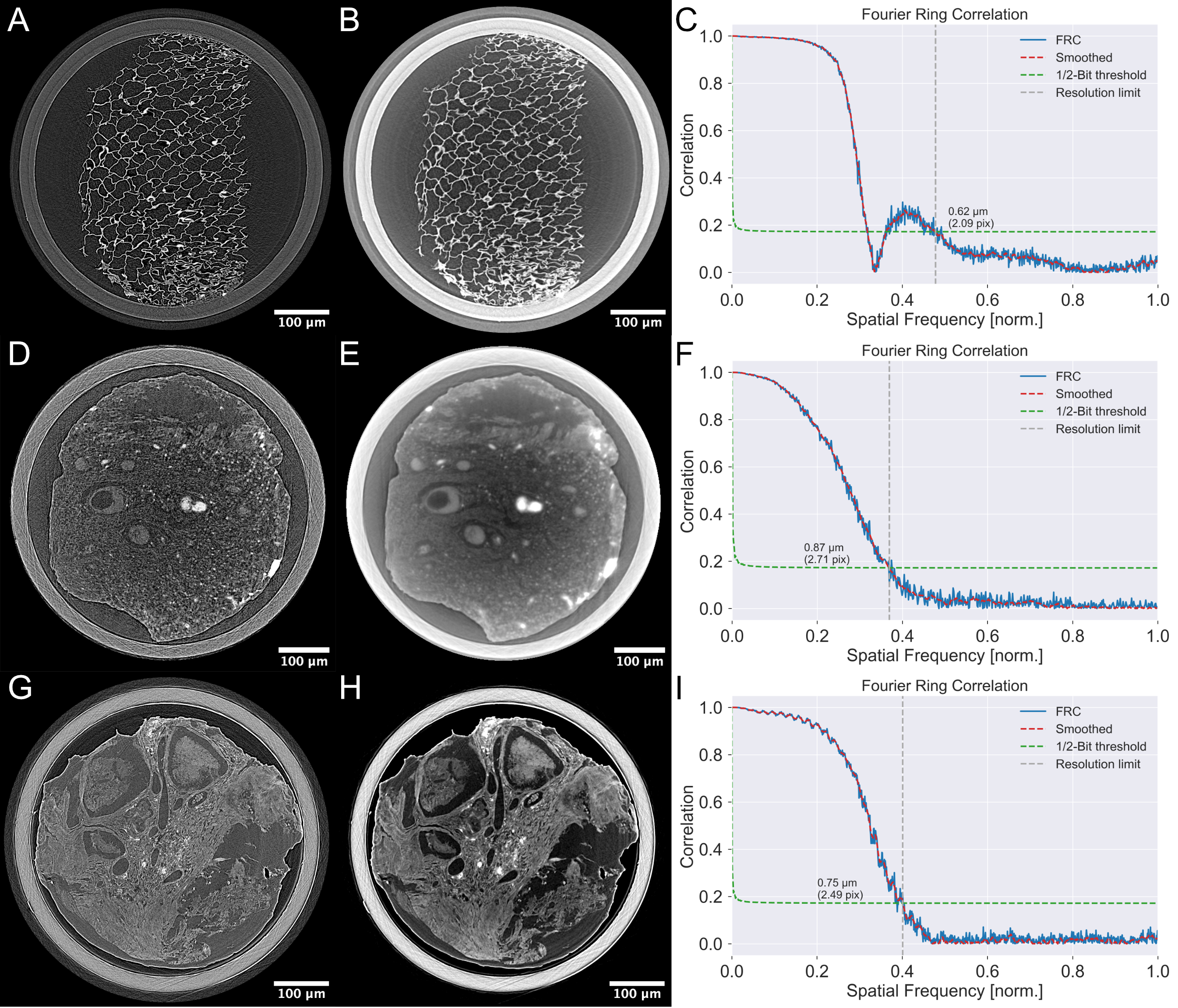}
    \caption{CT resolution using the GSense detector. (A) Reconstructed slice with absorption contrast of a 0.5~mm piece of cork with a voxel size of 297~nm. (B) Reconstructed slice with phase-retrieval applied. (C) Resolution determined by Fourier Ring Correlation. (D) Reconstructed slice with absorption contrast of a 0.5~mm punch fragment from a cow lung with a voxel size of 321~nm. (E) Reconstructed slice with phase-retrieval applied. (F) Resolution determined by Fourier Ring Correlation. (G) Reconstructed slice with absorption contrast of a 0.5~mm punch fragment from an atherosclerotic plaque with a voxel size of 298~nm. (H) Reconstructed slice with phase-retrieval applied. (I) Resolution determined by Fourier Ring Correlation.}
    \label{fig:pscam_resolution}
\end{figure}

% 3D resolution (PSCAM)
When using the GSense detector, the resolution of the three scans (Figure \ref{fig:pscam_resolution}) was found to be around 2 -- 3$\times p_{\text{eff}}$ corresponding to the expected PSF stated by the manufacturer. Since the results of the two scans appear consistent with each other, one could assume the cork scan with the Eiger detector shows in fact uncorrected tilts or drifts affecting the resolution.\\

% Figure: PSCAM FRC

\subsection{Contrast and noise}
% SNR
Evaluating the SNR (Equation \ref{eq:snr}) of the two detectors (Table \ref{tab:snr_cnr}) shows that concerning reconstructions without phase-retrieval, the SNR cannot be accurately measured with all measurements yielding around 49 dB. When applying phase-retrieval, the SNR yields a more reasonable number, possibly due to the smoothing, which is caused by phase retrieval. From a visual comparison, the GSense reconstructions look considerably more noisy, as seen in Figure \ref{fig:pscam_resolution}A, D, and G, as compared to the Eiger reconstructions shown in Figure \ref{fig:eiger_resolution}B, E, and H. The used regions for SNR and CNR calculations are shown in Supplementary Figure \ref{fig:S1}. Similar regions of interest (ROI) were selected for both detectors. The ROI used on the cork sample was placed on a denser part of the sample. Two ROIs were used on the lung sample, the first ROI was set on a dye filled vessel and the second on the tissue close to the airway in the upper part of the slice. Also, for the plaque sample, two regions were selected, focusing in calcified and non-calcified tissue respectively. \\

% CNR
In contrast to the SNR, the CNR (Equation \ref{eq:cnr}) yields a more robust measure on noisy images. Two things can be observed: (i) applying phase-retrieval significantly improves the CNR as evident from the reconstructions (Panels B, E in Figures \ref{fig:eiger_resolution} and \ref{fig:pscam_resolution}) and (ii) the Eiger detector performs significantly better, as expected. Hence, the CNR appears to be an appropriate measure to judge the improvement through phase contrast when dealing with low absorbing samples and generally noisy images. \\

\begin{table}
    \centering
    \caption{Measured SNR and CNR on the cork, lung, and plaque samples with both detectors. Regions shown in Supplementary Figure \ref{fig:S1}.}
    \begin{tabular}{|l|c|c|c|c|}
        \hline
                            & GSense (abs.) & GSense (phase)    & Eiger (abs.)  & Eiger (phase) \\
        \hline
         SNR: Cork          & 49.34 dB      & 26.13 dB          & 49.37 dB      & 28.42 dB \\
         SNR: Lung (ROI 1)  & 49.40 dB      & 22.83 dB          & 48.91 dB      & 35.10 dB \\
         SNR: Lung (ROI 2)  & 49.39 dB      & 23.08 dB          & 48.95 dB      & 34.68 dB \\
         SNR: Plaque (ROI 1)& 49.93 dB      & 30.48 dB          & 49.72 dB      & 26.89 dB \\
         SNR: Plaque (ROI 2)& 49.94 dB      & 30.28 dB          & 49.72 dB      & 26.88 dB \\
         \hline
         CNR: Cork          & 1.91          & 11.99             & 0.78          & 17.56 \\
         CNR: Lung (ROI 1)  & 0.62          & 1.14              & 3.65          & 12.52 \\
         CNR: Lung (ROI 2)  & 0.72          & 1.68              & 1.73          & 6.14 \\
         CNR: Plaque (ROI 1)& 3.09          & 10.91             & 4.3           & 8.06 \\
         CNR: Plaque (ROI 2)& 2.56          & 9.32              & 3.91          & 7.91 \\
         \hline
    \end{tabular}
    \label{tab:snr_cnr}
\end{table}

\section{Conclusions}
% setup performance
In this paper, we have presented the design and implementation of an x-ray nano-CT setup for biological samples capable of reaching sub-micron resolution and satisfying contrast. Data processing, including alignment, drift correction, and tilt correction, were described. Further, we have shown that the contrast can be increased by utilising propagation-based phase contrast while retaining the resolution in 3D through incorporation of unsharp masking into the phase-retrieval workflow. \\

% scintillator vs photon counting
The presented experiments show that, for biological applications with low contrast samples and low flux sources, photon counting detectors are the preferred choice. Resolutions close to the ideal single-pixel PSF of the detector could be achieved. Due to their high efficiency around 10~keV, the main energy in the x-ray spectrum, the photon counting detector performed significantly better than the scintillator-based detector despite the much larger distances required due to its larger pixels. \\

% tissue applications
It has been shown that the setup is capable of producing CT scans that are somewhat comparable to synchrotron scans, all though at significantly longer exposure times and requiring the sample to be very close to the source, i.e. typically a biopsy punch. Typically, tissue samples are embedded into larger paraffin blocks on which ROI scans are performed at a synchrotron. A downside of using an x-ray CT setup, as described in this paper, is the required sample size. To achieve high resolution, the sample has to be as close as possible, while keeping the whole system as compact as possible. Many features in e.g. tissue samples do not require the theoretically maximum achievable resolution of the x-ray source at 150~nm, but rather a larger FOV at relaxed resolution. The main limiting factor in the current system is the detector size, a wider detector would allow to scan larger samples at higher resolution.  \\

\section*{Ethical Approval}
The local Swedish Ethical Review Authority approved the use of the human atherosclerotic plaque samples (diary number 472/2005). The bovine lung samples were obtained from a local slaughterhouse and approved for research use by the local Swedish authorities (Jordbruksverket) under diary number 6.7.18-16885/2021.

\section*{Acknowledgements}
This project has been supported by the Swedish Foundation for Strategic Research (SSF grant ID17-0097) and by the Swedish research council (VR grant 2022-04192). The project has received funding from the European Research Council (ERC) under the European Union’s Horizon 2020
research and innovation programme (Grant agreement No. 101089334). Further, the authors acknowledge infrastructure funding from the Faculty of Science at Lund University. 
IG acknowledges support from the Swedish Research Council (2019-01260; 2023-02368), Swedish Heart and Lung foundation, and the Swedish Research Council, Strategic Research Area Exodiab, Dnr 2009-1039 (as EXODIAB member).
Jesper Wallentin and Hanna Dierks (Synchrotron Radiation Research, Lund University) are acknowledged for input and discussions during the design of the setup. Daniel Nilsson, Daniel Larsson, and Tomi Tuohimaa (Excillum AB) are acknowledged for their help and guidance using the source.

\bibliographystyle{apalike}
\bibliography{bibliography.bib}

\newpage

\section*{Supplementary}
% reset figure labelling
\renewcommand\thefigure{S\arabic{figure}}    
\setcounter{figure}{0}  

\begin{figure}[h]
    \centering
    \includegraphics[width=\textwidth]{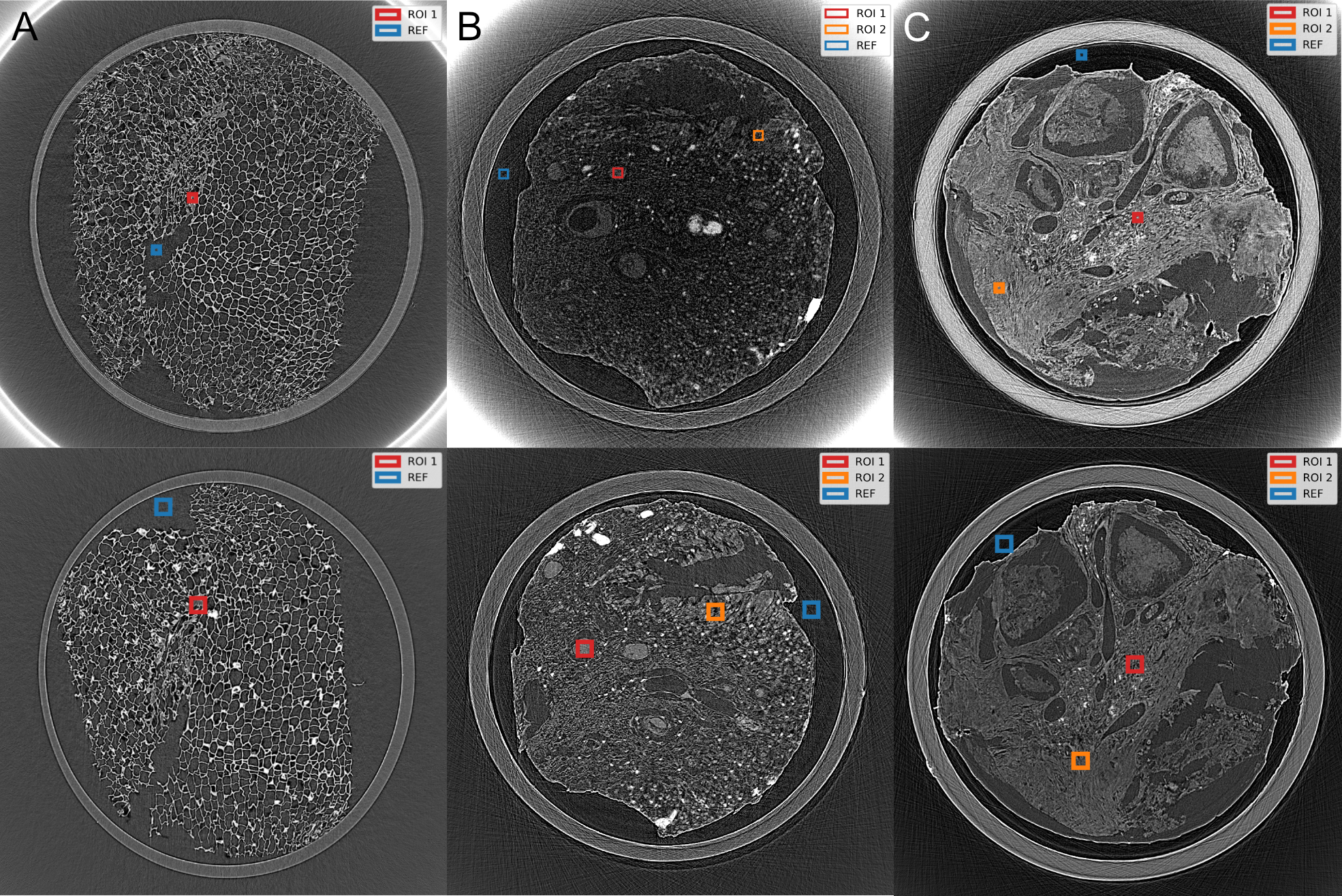}
    \caption{Regions used for SNR and CNR calculations. (A) Cork sample with the GSense and Eiger detector. (B) Lung sample with the GSense and Eiger detector. (C) Plaque sample with the GSense and Eiger detector.}
    \label{fig:S1}
\end{figure}

The selected regions for SNR and CNR calculations from the different reconstructed slices are shown in Figure \ref{fig:S1}. Background regions are marked in blue and are placed on regions containing air. Regions-of-interest (ROIs) are placed on the sample, indicated in red and orange. On the lung sample (Figure \ref{fig:S1}B), two regions are set: one on a dye filled vessel, and one on the tissue around an airway in the upper part of the image, respectively. On the plaque sample (Figure \ref{fig:S1}C), two regions are set on tissue of high and low density, repectively.

\begin{figure}[h]
    \centering
    \includegraphics[width=\textwidth]{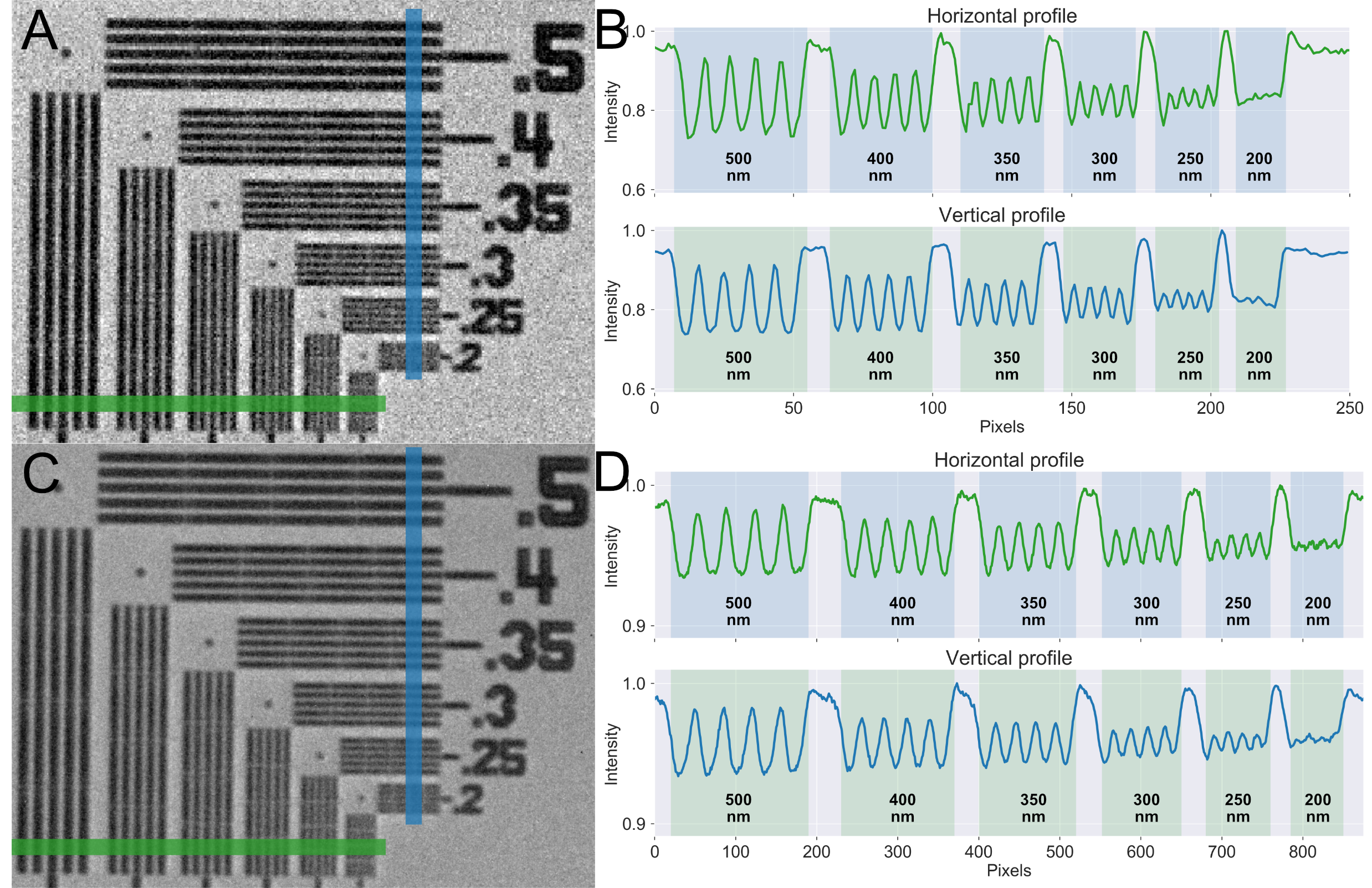}
    \caption{Horizontal and vertical line profiles through the 500 -- 200~nm features on the XRnanotech resolution chart. (A) Chart images with the Eiger detector at $p_{\text{eff}}~\approx$~150~nm. (B) Horizontal and vertical line profiles. (C) Chart imaged with the GSense detector at $p_{\text{eff}}~\approx$~50~nm. (D) Horizontal and vertical line profiles.}
    \label{fig:S2}
\end{figure}

The line profiles over the 500 -- 200~nm features on the XRnanotech chart, presented in Figure~\ref{fig:S2}, show that the 250~nm features can be resolved with both detectors in both directions with effective pixel sizes of approximately 50~nm and 150~nm for the GSense and Eiger detector respectively.

\end{document}